\documentclass[twocolumn,showpacs,showkeys,prd,superscriptaddress]{revtex4-2}
\usepackage{amsmath,amssymb,amsfonts,dsfont,mathrsfs,amsthm}
\usepackage{graphicx}
\usepackage{centernot}

\usepackage{xcolor}
\usepackage{natbib}
\usepackage{url}
\usepackage{hyperref}

\hypersetup{
    colorlinks=true,
    urlcolor=black,
    linkcolor=black,
    citecolor=black,
    linktocpage,
    pdfborder={0 0 0}
}

\usepackage{feynmf}
\usepackage{siunitx}
\usepackage{array}
\usepackage{ulem}
\usepackage{tikz}
\usepackage{braket}
\usepackage{epsfig}
\usepackage{epstopdf}
\usepackage{tabularx}
\usepackage{float}

\usepackage{booktabs} 

\usepackage{etoolbox}
\begin{document}

\title{Berezinskii-Kosterlitz-Thouless to BCS-like superconducting transition crossover driven by weak magnetic fields in ultra-thin NbN films}

\author{Meenakshi \surname{Sharma}}
\affiliation{School of Science and Technology, Physics Division, University of Camerino,
Via Madonna delle Carceri, 9, 62032 - Camerino (MC), Italy}
\author{Sergio \surname{Caprara}}
\affiliation{Dipartimento di Fisica, "Sapienza" Università di Roma (RM), Italy}
\author{Andrea \surname{Perali}}
\affiliation{School of Pharmacy, Physics Unit, University of Camerino,
Via Madonna delle Carceri, 9, 62032 - Camerino (MC), Italy}
\author{Surinder P. \surname{Singh}}
\affiliation{CSIR - National Physical Laboratory, Dr. K.S. Krishnan Marg, 110012 - New Delhi, India}
\author{Sandeep \surname{Singh}}
\affiliation{CSIR - National Physical Laboratory, Dr. K.S. Krishnan Marg, 110012 - New Delhi, India}
\author{Matteo \surname{Fretto}}
\affiliation{Advanced Materials Metrology and Life Science Division, INRiM (Istituto Nazionale di Ricerca Metrologica), Strade delle Cacce 91, Torino}
\author{Natascia \surname{De Leo}}
\affiliation{Advanced Materials Metrology and Life Science Division, INRiM (Istituto Nazionale di Ricerca Metrologica), Strade delle Cacce 91, Torino}
\author{Nicola \surname{Pinto}}
\affiliation{School of Science and Technology, Physics Division, University of Camerino,
Via Madonna delle Carceri, 9, 62032 - Camerino (MC), Italy}
\affiliation{Advanced Materials Metrology and Life Science Division, INRiM (Istituto Nazionale di Ricerca Metrologica), Strade delle Cacce 91, Torino}

\begin{abstract}
The Berezinskii-Kosterlitz-Thouless (BKT) transition in ultra-thin NbN films is investigated 
in the presence of weak perpendicular magnetic fields. A jump in the phase stiffness at the 
BKT transition is detected up to 5\,G, while the BKT features are smeared between 
5\,G and 50\,G, disappearing altogether at 100\,G, where conventional current-voltage behaviour  
is observed. Our findings demonstrate that weak magnetic fields, insignificant in bulk 
systems, deeply affect our ultra-thin system, promoting a crossover from Halperin-Nelson fluctuations 
to a BCS-like state with GL fluctuations, as the field increases. This 
behavior is related to field-induced free vortices that screen the vortex-antivortex 
interaction and smear the BKT transition.
\end{abstract}

\maketitle

In superconducting (SC) low-dimensional systems, the interplay between quantum coherence, interactions and topology can lead to unexpected emergent quantum behavior.
A clean two-dimensional (2D) superconductor undergoes a Berezinskii-Kosterlitz-Thouless (BKT) phase transition \cite{kosterlitz1973ordering,yong2013robustness,bartolf2010current,hu2020evidence,benfatto2009broadening,PhysRevB.100.064506,mondal2011role,Weitzel_PRL2023} that involves the proliferation of thermally activated vortex-antivortex pairs (VAPs) with an abrupt jump of the superfluid stiffness and the disruption of the quasi-long range ordered SC state, eventually promoting a pseudogap above the transition \cite{Franz1998,Marsiglio2015}.

The BKT physics has been observed in numerous ultra-thin SC films \cite{hsu1992superconducting,caprara2009paraconductivity,bombin2019berezinskii,zhang2008berezinskii,sharma2022complex,benfatto2009broadening}.
However, the effect of low magnetic fields 
($H\sim H_\mathrm{C1}$, $H_\mathrm{C1}$  
being the lower critical magnetic field) 
has not yet been investigated in detail, despite its importance 
for 2D SC systems, as mostly assessed in this work. The scarce available literature \cite{hsu1992superconducting,hu2020evidence,van1990superconducting,mironov2018charge} 
claims that the SC properties of 2D systems can 
undergo significant changes even at low $H$ fields, due to 
several factors, such as inhomogeneity, disorder, pinning mechanisms, 
occurrence of edge vortices and topological defects 
\cite{van1990superconducting,fisher1980flux}.
For a 3D superconducting system, an applied $H$ field of sizable intensity ($H > H_\mathrm{C1}$) shifts the
SC critical temperature, $T_\mathrm{C}$, to a lower value with a broadening of the SC transition due to the pair-breaking effects \cite{mathur1972lower,ashkin1984upper,chockalingam2008superconducting}. However, when the sample thickness, $d$, is reduced to $d\approx\xi$, several phenomena emerge, such as, e.g., the appearance of resistive tailing-like features, which have been reported by several groups \cite{hsu1992superconducting,sharma2022complex,joshi2018superconducting}, 
owing to their increased sample granularity \cite{lamura2002granularity, soldatenkova2021normal}, 
inhomogeneity and vortex pinning effects \cite{shapoval2011quantitative}.
It has also been argued that the value of 
$H_\mathrm{C1}$ vanishes for granular 2D SC systems  
\cite{il2008current,garland1987influence,gray1985resistance}, resulting 
into the generation of a large number of free vortices even at very low 
perpendicular fields. These vortices can easily penetrate into the material 
and closely interact with the VAPs, altering the dynamics of the BKT phenomenon.

In this letter, we present an experimental investigation of SC NbN ultra-thin films, a promising candidate for exploring the low-field regime \cite{sharma2022complex,benfatto2009broadening,bartolf2010current,venditti2019nonlinear}, aimed at achieving a comprehensive understanding of the fate of BKT transition and its evolution under weak magnetic fields ($3\div150$\,G).
Notably, NbN thin films exhibit a transition from 3D to 2D behaviour, when the thickness decreases below the scale of the SC coherence length, promoting the appearance of VAPs. Lowering 
 $T$, it becomes energetically favorable to bind the VAPs, resulting into a quasi-long range 2D SC state below $T_\mathrm{BKT}$.
To study this, we have measured the resistivity $\rho(T)$ of thin NbN films (see \hyperref[Suppl.Mat.]{Supplementary Material}) near $T_\mathrm{C}$, with changing the perpendicular $H$ field intensity. We have adopted the Aslamazov-Larkin (AL) Cooper pair fluctuation model \cite{aslamasov1968influence,larkin2005theory} and the Halperin-Nelson (HN) theory \cite{halperin1979resistive} as the two well-established paradigms to describe the gradual suppression of $\rho(T)$ when $T_\mathrm{C}$ is approached from above. We have analyzed the crossover from HN to 2D AL conductivity. 
Additionally, a critical analysis of the voltage-current ($V-I$) 
characteristics, measured by a pulsed mode technique (see \hyperref[Suppl.Mat.]{Supplementary Material}) to minimize sample heating artifacts \cite{Weitzel_PRL2023}, has allowed to assess the discontinuous jump in the superfluid stiffness with increasing $H$.\\
The analysis of $\rho(T)$ near $T_\mathrm{C}$, has evidenced a strong $H$  dependence, with a pronounced downward shift of $T_\mathrm{C}$ for $H$ as low as 3\,G, which persists up to $\approx 33$\,G (see Fig.\, S1 in \hyperref[Suppl.Mat.]{Supplementary Material}).
Above this threshold, the shift in $T_\mathrm{C}$ progressively decreases. This trend can be attributed to the occurrence of a sizeable density of vortices, introduced by the applied field. In fact, for a 10\,nm-thick NbN film, 
$H_\mathrm{C1} = 4\phi_0 \ln(w/\xi)/{w^2}\approx 0.3$\,G, where $\phi_0$ is the magnetic 
flux quantum and $w= 50$\,$\mu$m is the sample width \cite{gray1985resistance}. As 
a result, the induced vortex density $n_f=H/\phi_0$ varies from 
$\approx 1.45\times 10^7$\,cm$^{-2}$ to $\approx 7.25\times 10^8$\,cm$^{-2}$, for $H$ 
increasing from 3\,G to 150\,G. Even the lowest applied field is ten times larger than the calculated $H_\mathrm{C1}$, making it suitable for generating a sizable vortex density and resulting in a substantial reduction of $T_\mathrm{C}$.

These findings are consistent with previous investigations \cite{bartolf2010current} and numerical analysis 
of the interplay between the sample width and the coherence length  
\cite{mcnaughton2022causes}. 
A necessary condition $w > 4.4\,\xi(T)$ for vortex generation was proposed 
\cite{engel2008temperature}, and a phase diagram explaining the occurrence of vortices 
at very low $H$ intensities was obtained for large $w/\xi$ ratios
\cite{il2008current,mcnaughton2022causes}.
The tiny anomalies observed in the sample MO10 approaching $H=10$\,G, both in $T_\mathrm{C}$ and $\Delta T_\mathrm{C}$, have also been clearly detected for $d = 5$\,nm 
(MO5) \cite{sharma2022complex, joshi2020dissipation, constantino2018emergence}. 
\begin{figure}[htb]
\centering
\includegraphics[width= 8.6 cm]{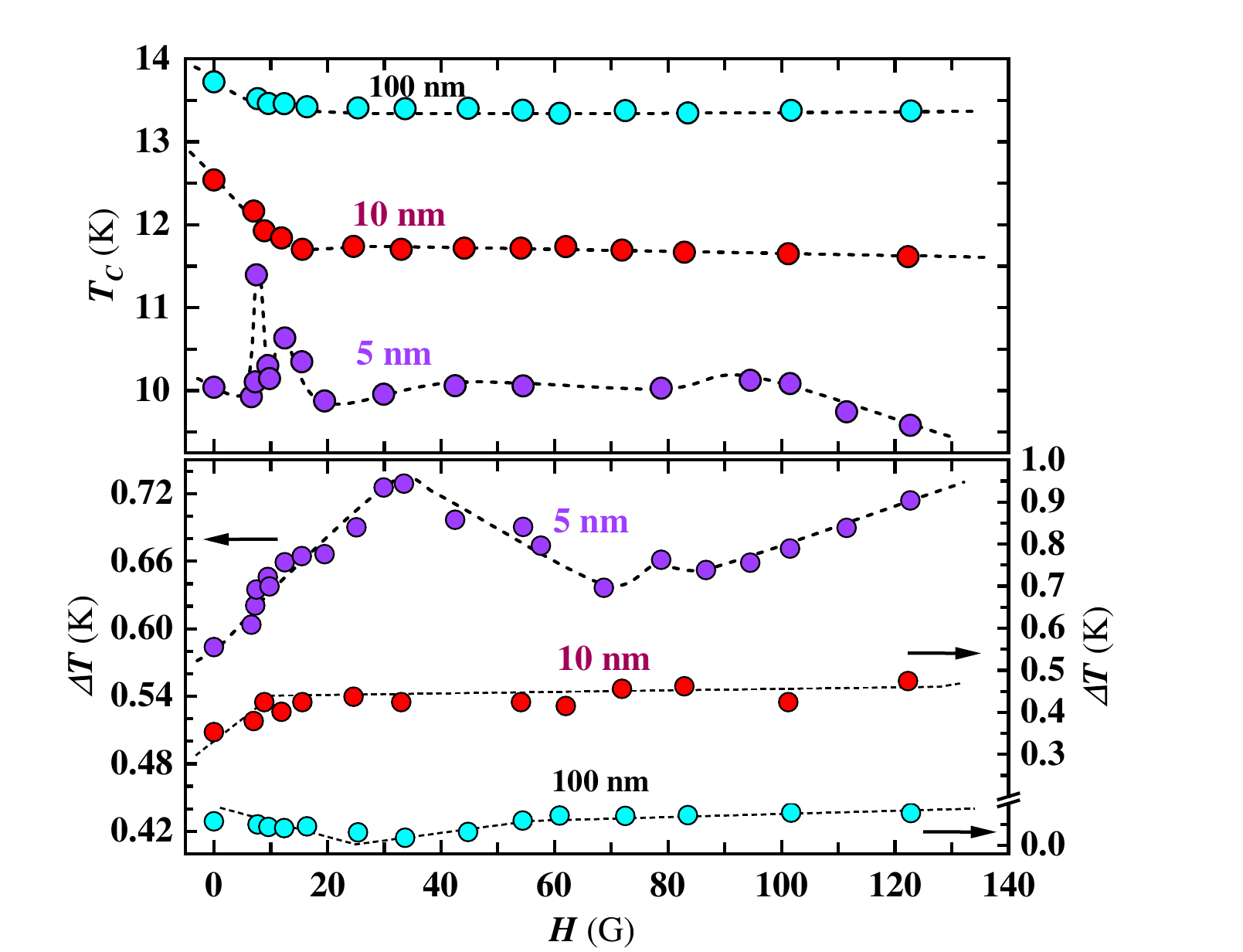}
\caption{Variation of $T_\mathrm{C}$ (a) and of the SC transition width (b) as a function of $H$ for NbN films of diverse thickness. Dash lines are guides for the eye.}
\label{Critical temperature}
\end{figure}
Remarkably, a resonance effect in $T_\mathrm{C}$ has been found in MO5 up to approximately 20\,G. Thereafter, a saturation effect persists up to $\approx 90$\,G, followed by a further decrease in $T_\mathrm{C}$ at $H > 90$\,G
(Fig.\,\ref{Critical temperature}a). 
Finally, at 100\,nm of thickness (MO100), the effect of $H$ on $T_\mathrm{C}$ is significantly reduced, with $T_\mathrm{C}$ shifting by a negligible amount ($\approx 0.2$\,K) up to 20\,G, then becoming constant (Fig.\,\ref{Critical temperature}a).
This is caused by the 2D-3D crossover when the thickness increases, with the gradual disappearance of the BKT physics that we detected in all the transport properties for this film.
Interestingly, a strong non-monotonic dependence of $\Delta T_\mathrm{C}$ on $H$ has been found at 5 nm (MO5), with an oscillatory $H$ dependence of $\Delta T_\mathrm{C}$ (Fig.\,\ref{Critical temperature}b). For thicknesses of 10\,nm (MO10) and 100\,nm (MO100), the variation in $\Delta T_\mathrm{C}$ occurs up to $H \simeq 15$\,G, 
with a very small overall variation in the thicker 
films (Fig.\,\ref{Critical temperature}b).
Such disparity in the SC behaviour of NbN films of different thickness at low $H$, is attributed to the inherent disorder that arises in systems whose 
thickness is comparable to $\xi$ ($\approx 5$\,nm). 
Increasing the film thickness towards $100$\,nm, 
fluctuation effects are supposed to be reduced due to the 2D-3D dimensional crossover \cite{Marsiglio2015} and lower expected 
disorder, as indeed observed in 
MO100 (Fig.\,\ref{Critical temperature}b).
However, in the case of ultra-thin films, the downward shift of $T_\mathrm{C}$ occurs in the low range of $H$ fields, suggesting a key role played by the first generated vortices to cause fluctuations and weakening of the VAPs interactions in the system.
Regarding the resonance of $\approx 2$\,K of $T_\mathrm{C}$ detected in MO5 for very tiny fields, its microscopic origin deserves further analysis. We argue that a complex interplay between disorder, coherence, and magnetic length scales can generate non linear effects in the vortex-antivortex interaction, leading to $T_\mathrm{C}$  amplifications. 
\begin{figure}[htb]
\centering
\includegraphics[width= 8.6 cm]{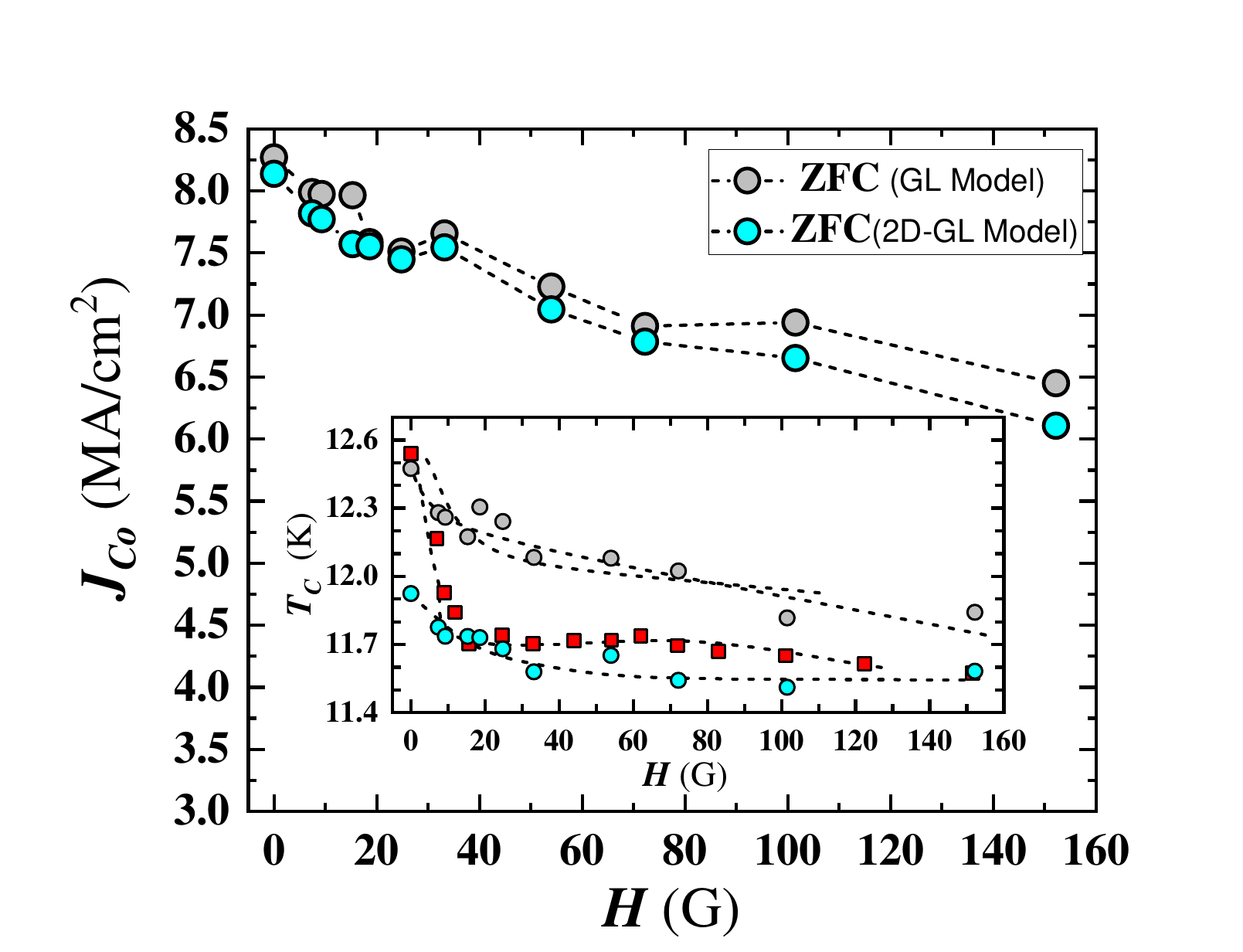}
\caption{Critical current density at 0\,K as a function of $H$ for MO10, under zero field cooled conditions, derived by a least square fitting with the 3D GL (grey circles) and the 2D GL equations 
(cyan circles). Inset: $H$ dependence of $T_\mathrm{C}$ derived by the least squares fitting of $\rho(T)$ curves (red squares) and of $V-I$ curves by 3D (grey circles) and 2D (cyan circles) GL equations. Dash lines are guides for the eye.} 
    \label{I-V}
\end{figure} 
To gain deeper insight on the above discussed phenomenology, we analyzed the $V-I$ characteristics under a perpendicular $H$ field. The critical current density at 
zero temperature, $J_\mathrm{C0}$, derived by the 3D and 2D GL equations fitting \cite{pinto2018dimensional} (see \hyperref[Suppl.Mat.]{Supplementary Material}), exhibits a similar reduction with increasing $H$ under zero field cooled (ZFC) measurements (Fig.\,\ref{I-V}), suggesting that $J_\mathrm{C0}$ is not sensitive to the dimensionality of the system. However, differences emerge considering $T_\mathrm{C}$. In fact, above 10\, G, $T_\mathrm{C}$ extracted from resistivity 
mimic the values derived by the 2D GL equation fit (Fig.\,\ref{I-V}, inset), confirming the 2D nature of our NbN films.
\begin{figure}[htb]
\centering
\includegraphics[width=0.9\linewidth]{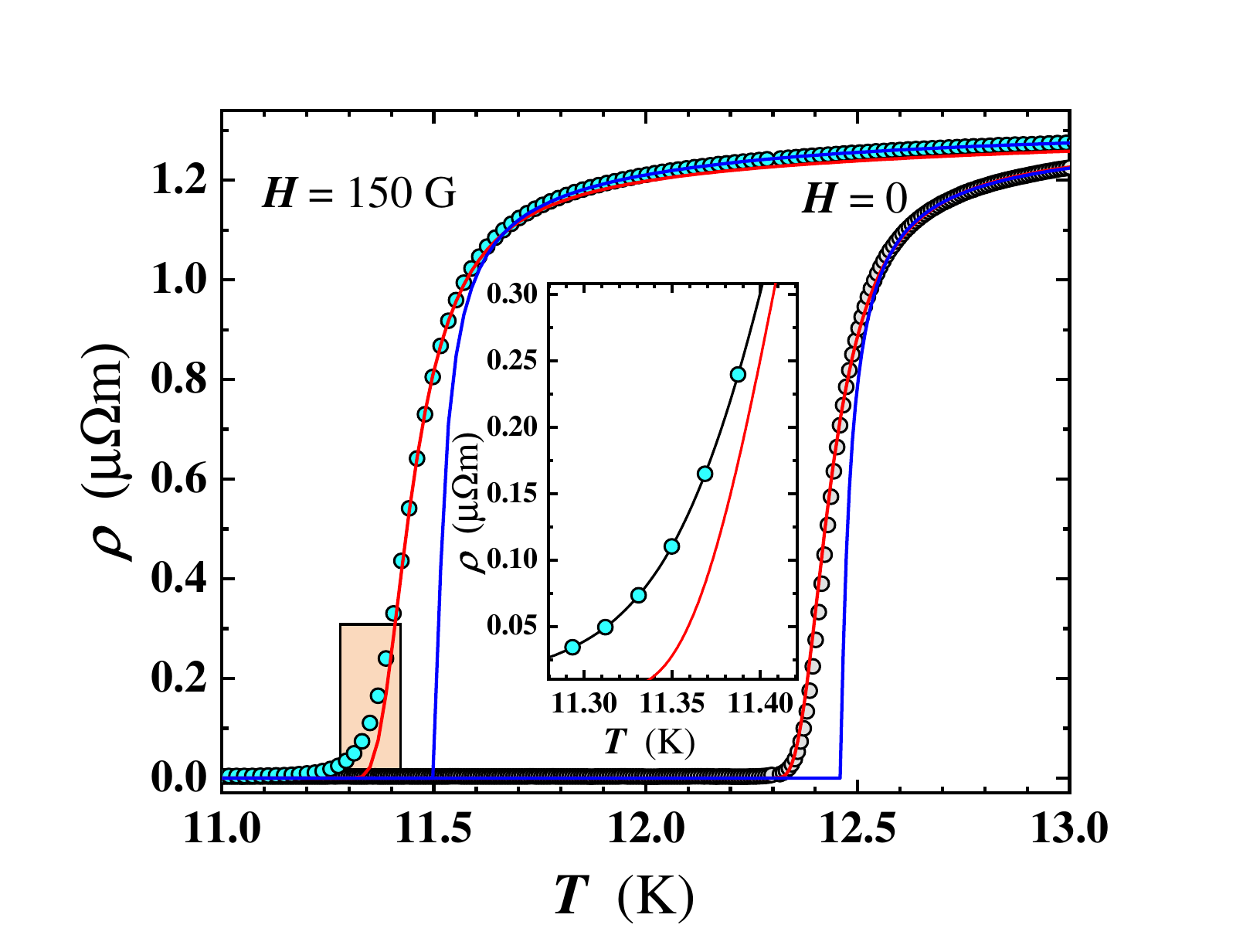}
    \caption{Temperature dependence of the resistivity for MO10, at $H = 0$\,G and $H = 150$\,G. Lines are least squares fitting by the HN equation [Eq.\,\eqref{Halperin-Nelson fluctuation conductivity}, red] and by the AL equation [Eq.\,\eqref{aslamazov-larkin}, blue].
    Smearing of the BKT phenomenon occurs at higher $H$ fields. Inset: magnification of the orange highlighted region.}
    \label{Resistance BKT}
\end{figure} 

To assess the occurrence of BKT, we have studied the $T$ dependence of $\rho$(T), close to the SC transition, revealing interesting features. The normal state resistivity is fitted at the 
highest field, $H=150$\,G, by $\rho_{N}(T)=A+B/T+C/T^2+D/T^3$.
This protocol of using the same normal-state resistivity at all magnetic fields reduces the number of fitting parameters, at the expenses of having slightly less accurate fits at low magnetic fields. 
The resulting least squares fitted parameters are: $A = 1.65 \times 10^{-6}$\,$\Omega\cdot$m, 
$B = -1.73 \times 10^{-5}$\,$\Omega\cdot$m$\cdot$K, $C = 3.02
\times 10^{-4}$\,$\Omega\cdot$m$\cdot$K$^{2}$, $D = -1.83 \times 10^{-3}$ $\Omega\cdot$m$\cdot$K$^{3}$. 
A relevant quantity is the $T$ dependant excess conductivity due 
to SC fluctuations, $\Delta\sigma = \sigma(T)-\sigma_N(T)$, 
where $\sigma_N(T)=[\rho_{N}(T)]^{-1}$ is the normal state conductivity.
The experimental paraconductivity 
is defined in term of the measured resistivity as 
$\Delta\sigma_\mathrm{exp}(T)=[\rho(T)]^{-1}-[\rho_N(T)]^{-1}$.
The AL paraconductivity is
\begin{equation}
    \Delta\sigma_\mathrm{{AL}}(T)=\frac{1521}{s\sinh{\left(\epsilon/s\right)}}
    \label{aslamazov-larkin}
\end{equation}
where 1521\,($\Omega\cdot$m)$^{-1}=e^2/(16\hbar d)$ is the prefactor for $d = 10$\,nm (MO10). The high-$T$ dimensionless cutoff $s$, possibly related to the pseudo gap \cite{caprara2005extended}, and $T_\mathrm{{AL}}$ are used as a fitting parameters, 
with $\epsilon = \ln\left(T/T_\mathrm{{AL}}\right)$.\\
The HN paraconductivity is
\begin{equation}
    \Delta\sigma_\text{HN}(T)=[\rho_N(T)]^{-1}a_\text{HN}\sinh\left[{\frac{b_\text{HN}
    \left(T_\mathrm{{HN}}\right)^{1/2}}{\left(T-T_\mathrm{{HN}}\right)^{1/2}}}\right]
    \label{Halperin-Nelson fluctuation conductivity}
\end{equation}
where the temperature $T_\mathrm{{HN}}$ and the dimensionless parameters 
$a_\text{HN}$ and $b_\text{HN}$ are taken as fitting parameters.
To plot the HN and AL paraconductivity in the same plot, that is a function of $\epsilon=\ln(T/T_\mathrm{{AL}})$, 
we use the identity $T-T_\mathrm{{HN}}=T_\mathrm{{AL}}\,\text{e}^{\epsilon}-T_\mathrm{{HN}}$. 
\begin{table}[H]
\caption{Parameters derived by the least square fitting of the sheet resistance data of film MO10, by using the 
Cooper-pair fluctuation models [i.e., Eq.\,(\ref{aslamazov-larkin}) and 
Eq.\,(\ref{Halperin-Nelson fluctuation conductivity})] for some selected $H$ intensities.} 
\vspace{0.25cm}
\begin{center}
\begin{tabular}{@{}|l|l|l|l|l|l| @{}}
\hline
H (G)& $T_\mathrm{{AL}}$ (K)& $~s$ &~$a_\text{HN}$&~$b_\text{HN}$&$T_\mathrm{{HN}}$ (K)\\
\hline

0 & 12.46 & ~0.075~ & ~0.0050~ & ~0.68~ & 12.285 \\
7 & 12.07 & ~0.057~ & ~0.0048~ & ~0.68~ & 11.88 \\
50 & 11.62 & ~0.029~ & ~0.0048~  & ~0.72~ & 11.42 \\
100 & 11.55 & ~0.032~ &  ~0.0048~ & ~0.72~ & 11.35 \\
\hline
\end{tabular}
\end{center}
\label{Table}
\end{table}

Except at $H = 0$, $a_\text{HN}$ is constant while $b_\text{HN}$ exhibit a minor dependence on $H$ (see Tab. I). Regarding $T_\mathrm{{AL}}$ and $T_\mathrm{{HN}}$ both monotonically lower with $H$, while $T_\mathrm{{AL}}-T_\mathrm{{HN}}\approx 0.2$\,K in the entire $H$ range (Tab. I).

\begin{figure}[htb]
    \centering
    \includegraphics[width=1.0\linewidth]{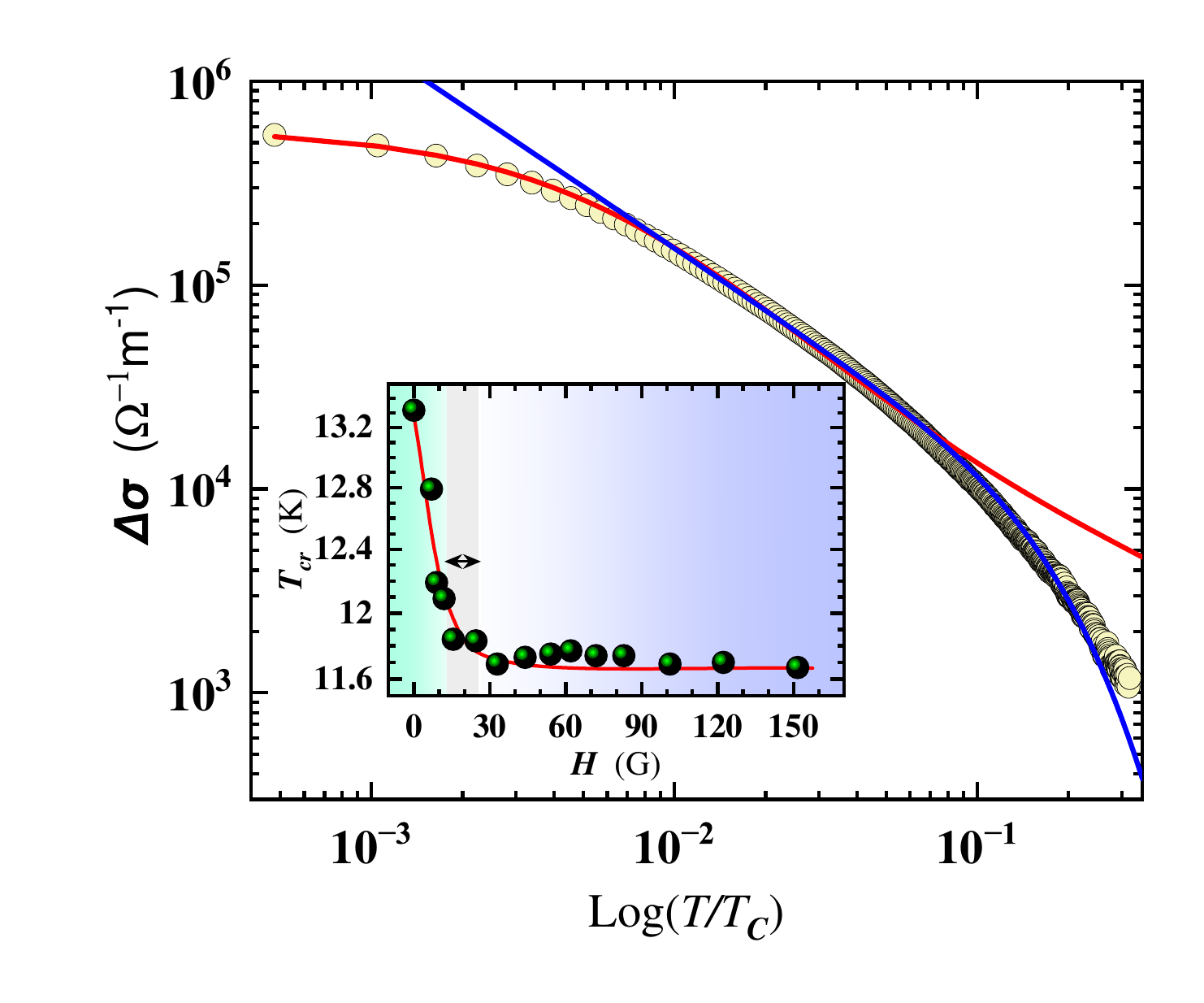}
    \caption{a) Paraconductivity of MO10 as a function of the reduced temperature, under an applied field of $H=0$\,G. Lines are least squares fitting derived by the AL (blue, Eq.\,\ref{aslamazov-larkin}) and HN (red, Eq.\,\ref{Halperin-Nelson fluctuation conductivity}) equations. Inset. $H$-field dependence of the crossover temperature, $T_{cr}$, defined as the crossing point of the AL and HN least square fitting curves by Eqs.\,
    (\ref{aslamazov-larkin}) and (\ref{Halperin-Nelson fluctuation conductivity}). The transition region from the BKT (pale green) to the BCS-like (pale blue) physics has been highlighted by a grey filled area ($\approx 15\div20$ G). The red line is a guide for the eye.}
    \label{paraconductivity}
\end{figure}

For $H = 0$, $\rho(T)$ is well fitted by 
Eq.\,(\ref{Halperin-Nelson fluctuation conductivity}), showing a clear manifestation of the BKT behaviour. However, under an applied $H$, Eq.\,(\ref{Halperin-Nelson fluctuation conductivity}) gradually fails to fit the data, both close $T_\mathrm{C}$, where a magnetic-field-driven foot develops, and further away from the transition, where the data are much better described by the AL theory
(Eq.\,\ref{aslamazov-larkin}, Fig.\,\ref{Resistance BKT}). We point out that there was no way to significantly improve the fitting by considering a Maki-Thomson contribution, presumably because this is suppressed by strong dephasing \cite{carbillet2016confinement}.
This gradual suppression of the BKT physics provides a clear evidence for the contribution of $H$-induced vortices, even at extremely weak magnetic fields, significantly screening the logarithmic interaction between VAPs, thereby affecting the dynamics of the interaction and effectively suppressing the BKT transition. 
However, this smearing of BKT physics does not significantly impact the SC state, that remains preserved, particularly in the weak field regime, allowing for a selective control of fluctuation effects.\\
To study the crossover in more detail, we analyze the paraconductivity of the films exhibiting the BKT effect. In the absence of any applied $H$ field, the HN approach, Eq.\,(\ref{Halperin-Nelson fluctuation conductivity}), succeeded to fit 
the fluctuation conductivity of a thin NbN film (MO10), in the region following the SC transition, where the BKT mechanism 
persists (Fig.\,\ref{paraconductivity}). 
In presence of $H$, the HN equation starts to deviate significantly from the experimental data, with a gradual crossover from the HN to the Ginzburg-Landau (GL) description, as evidenced by the progressive decrease in the crossover temperature, $T_{cr}$ (Fig.\,\ref{paraconductivity}, inset), defined as the temperature at which the crossover between the HN and the AL fitting functions of the PAR takes place. Therefore, the value of $T_{cr}$ enabled us to identify the $H$ field-induced crossover from BKT physics to GL physics (Fig.\,\ref{paraconductivity}, inset). These findings align with the observations made regarding the behavior of  
$T_\mathrm{C}$ for MO10 (see Fig.\,S1 in \hyperref[Suppl.Mat.]{Supplementary Material}).
In presence of $H$, we detect a progressive lowering of  $T_{cr}$, allowing to identify the $H$-field-driven crossover at $H\approx 15\div 20$ G from the BKT to the GL physics (Fig.\,\ref{paraconductivity}, inset), in agreement with  findings of $T_\mathrm{C}$ behaviour of sample MO10, (see Fig.\,S1 in \hyperref[Suppl.Mat.]{Supplementary Material}).

\begin{figure}[htb]
    \centering
    \includegraphics[width=1.0\linewidth]{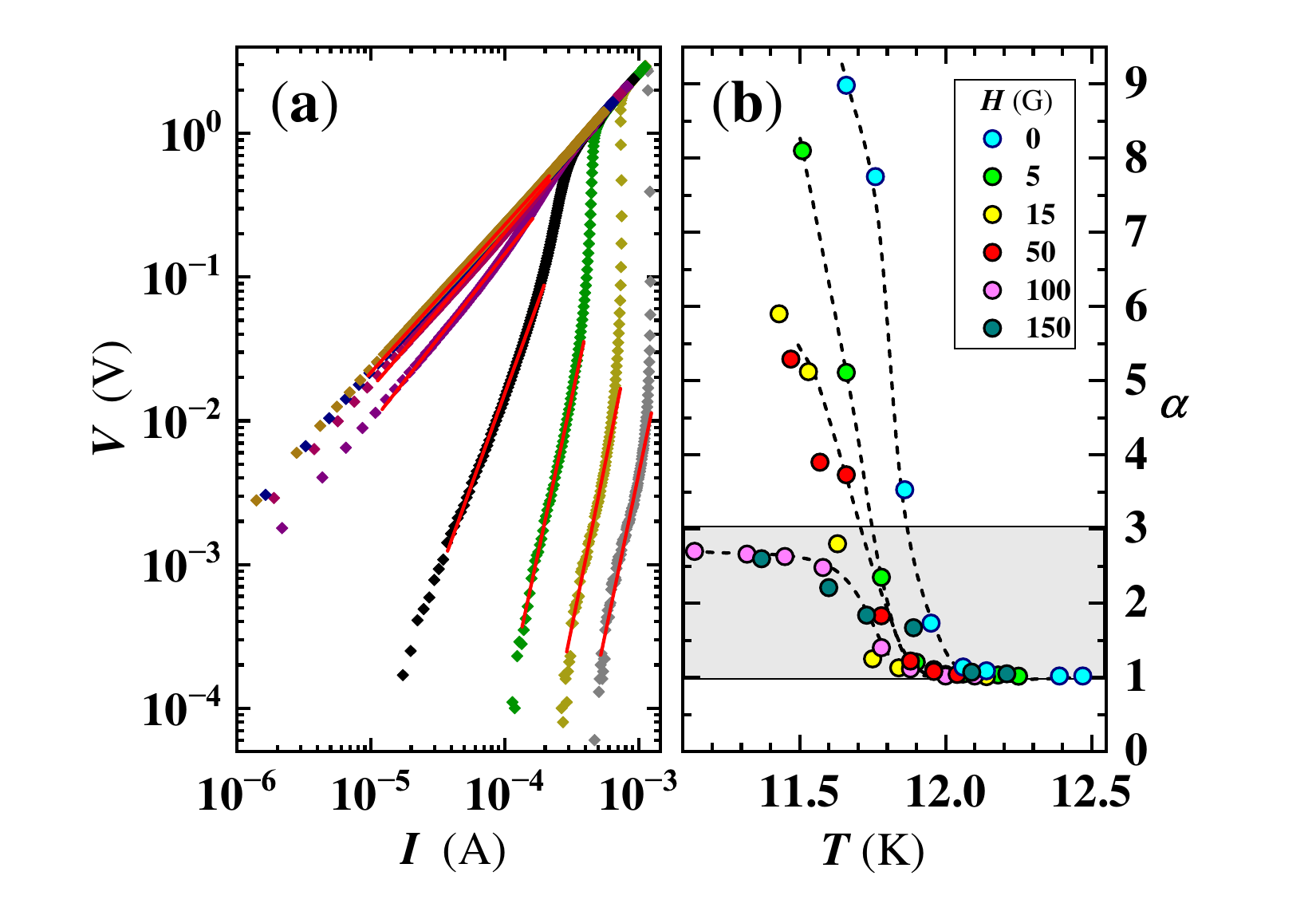}
    \caption{a) $V-I$ curves of MO10, at several $T$ close to $T_{\mathrm C}$, under an applied field of $H = 5$\,G. Red lines are least square fitting by the relation $V\propto I(T)^{\alpha(T)}$ of curves measured at (from right to left): 11.30\,K, 11.43\,K, 11.53\,K, 11.63\,K, 11.75\,K, 11.84\,K, 12.0\,K, and 12.14\,K. b) Temperature dependence of the fitted $\alpha$ exponent for several $H$ field intensities.}
    \label{I-V&alpha}
\end{figure}

As previously stated, the presence of non-linear behavior in the $V-I$ 
characteristics arising from the discontinuous jump in the superfluid stiffness, $J_{S}$, serves as a distinctive feature of the BKT transition. Therefore, it is interesting to assess its occurrence in the presence of a magnetic field. Typically, the discontinuous jump in $J_{S}$ is determined by two distinct approaches: 
direct measurement of the inverse penetration depth \cite{venditti2019nonlinear} and 
indirect measurement of the exponent from the non-linear $V-I$ characteristics 
near $T_\mathrm{C}$. Both approaches result into similar outcomes, as investigated by 
Venditti et {\it al.} \cite{venditti2019nonlinear}.   
Here, we follow the latter. The power law dependence is given by $V \propto \left[I(T)\right]^{\alpha(T)}$, where 
$\alpha(T) = 1+\pi\,J_{S}(T)/T$. 
We determined the exponent $\alpha(T)$ from the slope of the $V-I$ characteristics, measured at several temperatures close to $T_C$, under various $H$ fields (Fig.~\ref{I-V&alpha}a).
\begin{figure}[htb]
    \centering
    \includegraphics[width=1.0\linewidth]{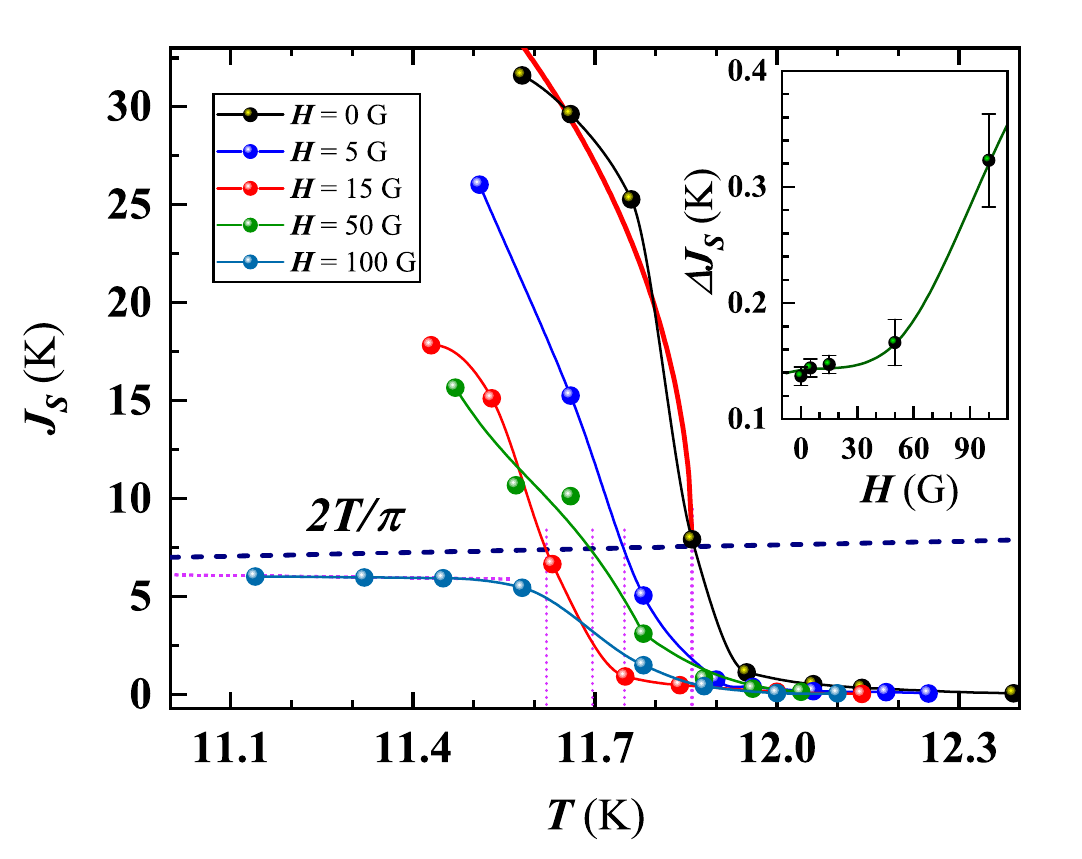}
    \caption{Temperature dependence of the SC stiffness of MO10, derived from $\alpha$ values at several $H$ field intensities (see 
    Fig.\,\ref{I-V&alpha}b). Experimental values at $H = 0$ G have been fitted by the Eq.(\ref{Stiffness}) for $T\leq T_{BKT} = 11.86$ K, resulting by the intersection of the $2T/\pi$ (blue broken) with $J_S$ interpolating lines (black). The fitted parameter of Eq.(\ref{Stiffness}) is $A = 10.4\pm0.6$ K$^{-1/2}$. Inset: field dependence of $J_S$ width, estimated as the $T$ difference of the 10$\%$ and 90$\%$ of the $J_S$ value, assumed as the intercept of the $2T/\pi$ line with the curve interpolating the experimental behavior. The error bars represent the uncertainty in the $T$ difference related to the choice of the curve interpolating data. The solid ine is a guide for the eye.}
\label{Js}
\end{figure}
We found a sizable $H$ dependence of $\alpha(T)$. At intensities as low as 5\,G, 
$\alpha$ is downward shifted and starts to lower its slope (Fig.~\ref{I-V&alpha}b). The shift continues up to 15 G, while at $H>50$\,G, a sizable lowering of the slope starts to occur, together with a discontinuous change in $\alpha$ (Fig.\,\ref{I-V&alpha}b). At 100\,G, increasing $T$ up to 11.6\,K, $\alpha$ saturates at a value of $\approx 2.5$, to finally converge to $\alpha = 1$ at $T\approx 12$\,K.
As it is known, at $T=T_\mathrm{BKT}$, $J_S$ is expected to jump discontinuously, from $J_S(T^{-}_\mathrm{BKT}) = 2T_\mathrm{BKT}/\pi$ to $J_S(T^{+}_\mathrm{BKT})=0$, giving $\alpha(T^-_\mathrm{BKT}) = 3$ and $\alpha(T^+_\mathrm{BKT}) = 1 $, respectively.

To gain more insight into the smearing of the BKT transition with increasing $H$, we have derived $J_S(T)$ close to the SC transition (Fig.\,\ref{Js}) from the extracted $\alpha$ values by using the relation $J_S(T) =\left[\alpha(T)-1\right]T/\pi$ \cite{venditti2019nonlinear}. In the case of $H=0$, $J_S$ shows quite an abrupt jump at $T_\mathrm{BKT}$ and its $T$ dependence before the transition has been fitted by the functional dependence of the renormalization group equations given by: 
\begin{equation}
J_S(T)=2T/\pi+4AT\sqrt{\left(T_\mathrm{BKT}-T\right)}/\pi^2
    \label{Stiffness}
\end{equation}
where the fitted parameter is $A = 10.4\pm0.6$ K$^{-1/2}$.
The agreement between experiment and theory is satisfactory (Fig.\,\ref{Js}), pointing toward an accurate BKT description of the superconducting transition at zero field. Increasing $H$, the fit fails and the sharpness of the jump starts to smear. To account for this smearing, we have estimated the width of $J_S$ ($\Delta J_S$) of the jump starting from the intersection with the universal line $2T/\pi$. The $\Delta J_S(H)$ trend further confirms the progressive smearing of the BKT physics increasing $H$ (Fig.\,\ref{Js}, inset).

To conclude, distinct features of the BKT transition in NbN ultra-thin films have evidenced a high sensitivity to perpendicular weak magnetic fields. Our experimental findings show that the BKT phenomenon survives up to $H = 5$ G, while increasing $H$ to $\approx 15\div 20$ G, causes a crossover from BKT physics to BCS-like SC transition. This is demonstrated by the failure of the Halperin-Nelson description of the fluctuation conductivity close to the BKT transition with an increasing $H$ and a corresponding improved convergence of the Aslamazov-Larkin equation with the experimental data, highlighting a GL fluctuation behavior on top of a BCS state. 
Above $H = 100$ G, any signatures of the BKT phenomenon are completely washed away. We correlate this behavior with the impact of field-induced vortices that screen and weaken the vortex-antivortex interaction, promoting a tunable crossover from BKT to BCS superconductivity. 
This study indicates the key role played by weak magnetic fields in 2D superconducting systems, otherwise negligible in 3D, that should be accounted for when designing integrated superconducting circuits. Our results suggest that similar effects could be found in analogous superfluid systems realized with fermionic atoms confined in 2D pancakes under slow rotations, a new field of investigation of BKT in 2D rotating superfluids.\\

{\it Acknowledgements.} We thank Giulia Venditti for fruitful discussions. 
Part of the experimental activity has been carried out at QR Lab - Micro\&Nanolaboratories, INRiM.
This work has been partially supported by PNRR MUR Project No. PE0000023-NQSTI
and by the ‘Ateneo Research Projects’ of the University of Rome Sapienza:  ‘Competing phases and non-equilibrium phenomena in low-dimensional systems with microscopic disorder and nanoscale inhomogeneities’ (n. RM12117A4A7FD11B), ‘Models and theories from anomalous diffusion to strange-metal behavior’ (n. RM12218162CF9D05), `Non-conventional aspects for transport phenomena and non-equilibrium statistical mechanics' (n. RM123188E830D258).
    

\clearpage
\section{Supplementary Material}
\label{Suppl.Mat.}
\subsection{Film Deposition}
Niobium nitride (NbN) films were deposited by using reactive DC magnetron sputtering from Excel Instruments, operational at the CSIR-National Physical Laboratory (NPL), India. The process utilized a 2-inch niobium target with a purity of 99.95$\%$, positioned 8.5 cm away from the substrate. The sputtering atmosphere comprised ultra-high purity (99.999$\%$) N$_2$ and Ar gases. The deposition process was carried out on four distinct substrate types: MgO, Si, and Al$_2$O$_3$ both r-and c-cut. Prior to deposition, each substrate was cleaned using acetone and isopropyl alcohol (IPA) to ensure surface purity and optimize film adhesion.
The deposition chamber was evacuated to a pressure of $10^{-7}$ mbar using a turbo-molecular pump. Simultaneously, the chamber's heater was brought up to the target temperature of 600 \(^\circ\)C. This temperature ramp was controlled through a pre-set program (temperature and time) using a PID controller to ensure precise thermal regulation. Once the desired temperature was achieved, the N$_2$ and Ar gases were injected into the chamber at a ratio of 1:6 and the pressure of the chamber was maintained at $7\times 10^{-3}$ mbar, where a self-sustain plasma formed at 200 W of DC power supply. After the plasma formation, a pre-sputtering of the target was done for $\approx$ 2 min, then followed by deposition of NbN. The deposition rate comes out to be 0.3 nm/s and the film thickness was measured by using a stylus profilometer (NanoMap 500ES).\\
\subsection{Device Fabrication}
\begin{figure}[ht]
    \centering
    \includegraphics[width=0.45\textwidth]{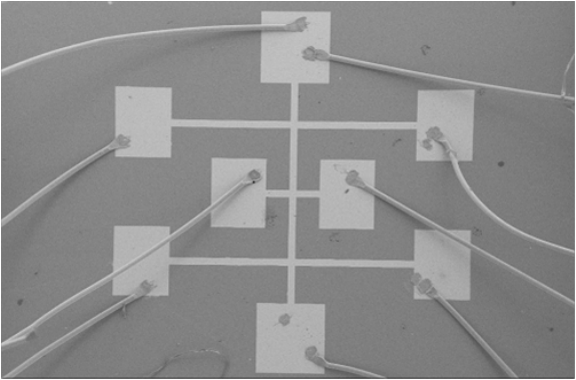}
    \caption{Scanning Electron Microscopy (SEM) image showcasing the Hall bar geometry of a 10 nm NbN thin film deposited on a MgO substrate. Pads at the center, both top and bottom, have been used to source the current while two out of three lateral pads, located on the same side, to measure the voltage drop. The photo also shows bonded aluminum wires used to connect the Hall bar to the pads of a PCB, this last allowing connection of the device to measuring apparatus (see Figure~\ref{PCB_COPPER_DISC}).}
    \label{SEM_Hall_bar}
\end{figure}

\begin{figure}[ht]
    \centering
    \includegraphics[width=0.45\textwidth]{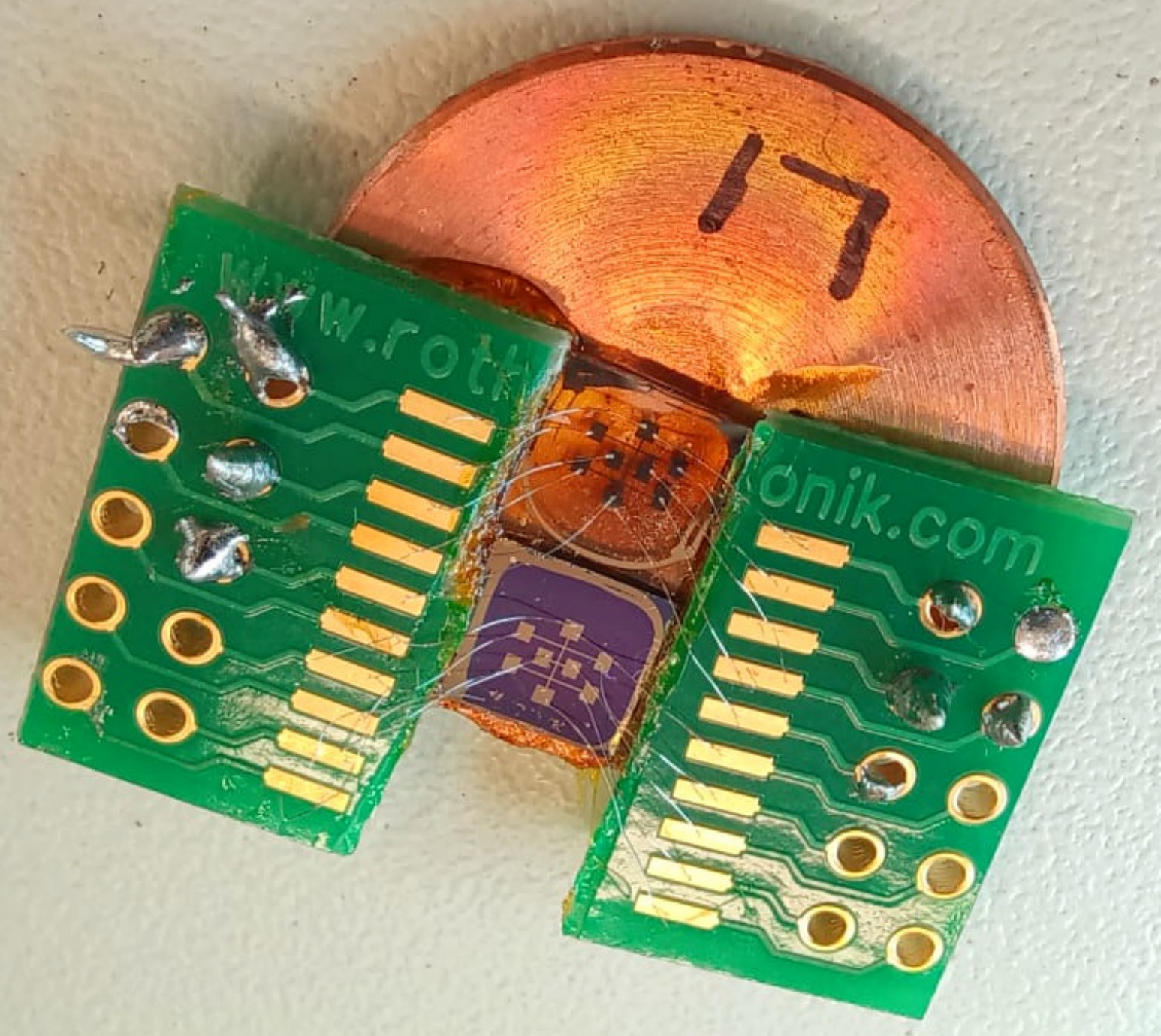}
    \caption{Examples of two NbN devices on Si (center, bottom) and MgO (center, top) substrates and of two PCBs, glued to the copper disc. The aluminum wires, bonded both to the Hall bar and PCB pads are clearly visible. The adopted solution while preventing Al wires detaching assure an optimal thermal contact of devices. The photo has been taken after the electrical characterization.}
    \label{PCB_COPPER_DISC}
\end{figure}

For the electrical measurements of transport properties of NbN films, a suitable Hall bar geometry has been designed and fabricated. We have chosen 8-contact geometry, designed by the LibreCAD software (see Figure~\ref{SEM_Hall_bar}). 
The fabrication steps involved spin coating of the photoresist (AZ 5214E) onto the deposited film by a spin coater operated at 4000 rpm. 
A soft bake was then performed at 110°C for 3 minutes, resulting in a nominal thickness of 1.4 $\mu$m for the photoresist layer. 
The pattern generation process involved a Karl Suss mask aligner (mod. MJB3) for optical lithography (line space resolution of 2.5 $\mu$m) followed by wet chemical developing.\\
Later, an etching step was used to define the final geometry of the NbN Hall bars. In particular, the etching process was carried out by using deep reactive ion etching (RIE: PlasmaPro 100 Cobra Inductively Coupled Plasma Etching System, Oxford Instruments). The applied RIE power have been 30 W (table power rf) and 500 W (ICP power rf). 
The fabrication parameters were optimized taking into account both film thickness and the substrate type. The best etching selectivity between the optical resist and the NbN thin film was obtained with a mixture of CF$_4$ and Ar with fluxes of 90 and 10 sccm, respectively, at an operating pressure of 50 mTorr. All the samples were cleaned with a light-oxygen plasma before the etching, to eliminate any organic or lithographic residuals.\\
The fabricated NbN devices and 1-2 small PCBs were glued (Oxford, GE Low Temperature Varnish) to a copper disk (see Figure~\ref{PCB_COPPER_DISC}). 
Finally, thin Al wires have been bonded to the Hall bar and PCB pads allowing electrical connection to the measuring equipment (Figures~\ref{SEM_Hall_bar} and ~\ref{PCB_COPPER_DISC}).\\
\subsection{Transport Measurements}
For the electrical characterization, copper disks with NbN devices and PCBs (Figure~\ref{PCB_COPPER_DISC}) have been thermally anchored to a copper cold finger by silver paste or Apiezon N grease.\\
Transport measurements have been carried out in a He closed-cycle (liquid-free) cryostat (Advanced Research System mod. DE210) equipped with two Si diode thermometers (Lakeshore mod. DT-670) one out of two calibrated (maximum uncertainty of 6.3 mK). Temperature of the second stage of the cryostat has been read by the uncalibrated Si diode wired to a controller Lakeshore mod. 332, while sample temperature has been read by one channel of a double channel source-measure unit (SMU, Keysight mod. B2912A). No change in the $T$ value has been detected under the perpendicularly applied  magnetic field, in the intensity range used in our study.\\
Resistivity, $\rho(T)$, and current-voltage ($I-V$), characteristics have been measured as a function of the temperature in the 4-contact geometry. For investigation of $\rho(T)$, the SMU has been operated either in dc or in pulsed mode. The latter (so called delta mode) is based on the suitable combination of three consecutive current pulses of alternate polarity, virtually able to minimize thermo-electric forces and offsets otherwise affecting the low voltage detected values \cite{Keithley_1}.\\
Current (1 $\mu$A$\div 10$ mA for $I-V$; typically $\leq 1 \mu$A for $\rho$) has been sourced along the Hall bar length and the voltage drop detected by using two out of three lateral contacts, located on the same side (Figure~\ref{SEM_Hall_bar}).\\
Electrical transport has been started from the lowest $T$ ($\approx 4.5$ K) reached by the system, without any thermal stabilization of the device, exploiting the high thermal inertia of the cryostat, resulting in effective iso-thermal conditions of measure.
In particular, the number of data points to average have been suitably chosen to keep the variation of the sample temperature as low as possible: $\lesssim 1$ mK, during data acquisition of the superconducting transition (fast measurement: $\approx 5$ values/s resulting each by the average of 30 readings); $\lesssim 10\div 20$ mK, for $I-V$ characteristics.\\
Regarding $I-V$ measurements, a fast pulsed sweep technique has been developed, exploiting a specific feature of the SMU. Current pulses of 1.1 ms of duration and of increasing intensity have been sourced to the device in both direction (i.e., up and down sweeps) collecting typically $150\div200$ points, for each $I-V$ curve. Between two consecutive pulses, the current is switched off for about 1 ms. This pulsed technique has been implemented to significantly reduce any artifact due to the heating effect of the investigated film.\\
An electromagnet Bruker mod. B-E15v has allowed to study the BKT phenomenon under low intensity applied magnetic fields, perpendicular to the film surface. 
$H$ field intensity and its uniformity within an area of about 80 cm$^2$ has been checked by an AlphaLab Inc. GM-2 gaussmeter having a sensitivity and a resolution of 0.1 G.\\
\begin{figure}[ht]
    \centering
    \includegraphics[width=0.45\textwidth]{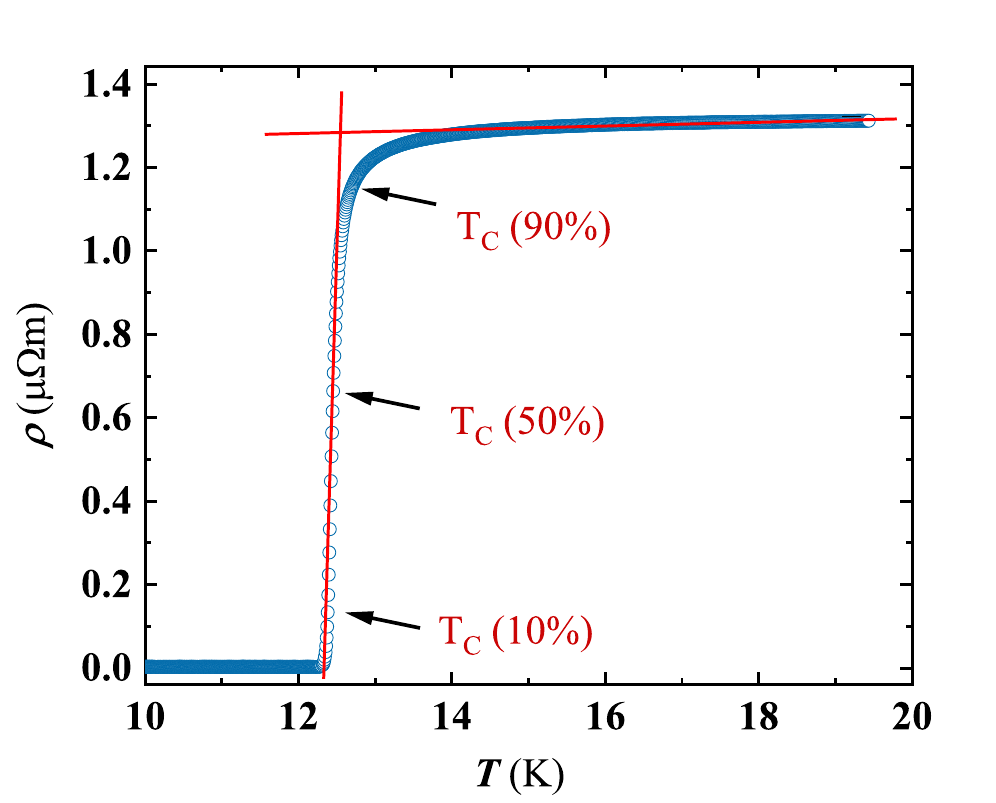}
    \caption{Temperature dependent resistivity of a 10 nm thick NbN film on MgO substrate measured at $H = 0$ G). The figure illustrates the procedure used to extract the  superconducting critical temperature ($T_\mathrm{C}$). For the definition see the text.}
    \label{Tc}
\end{figure}
\subsection{Superconducting Properties of NbN Films}
In this section, will be described methods, given definitions of the main SC parameters used in our experimental work and shown additional behaviors exhibited by investigated NbN thin films.\\
\subsubsection{$T_\mathrm{C}$ and $\Delta T_\mathrm{C}$}
The superconducting critical temperature, $T_\mathrm{C}$, of investigated NbN films has been defined as the midpoint between the two temperatures corresponding to the $10\%$ ($T_{10\%}$) and to the $90\%$ ($T_{90\%}$) of the normal state resistivity, upon film transition from the superconducting to the normal state, located on a resistivity plateau (see Figure~\ref{Tc}).The difference between the two temperatures has been defined as the width of the superconducting transition: $\Delta T_\mathrm{C} = T_{90\%} - T_{10\%}$.
\begin{figure}[htb]
\centering
\includegraphics[width=8.cm]{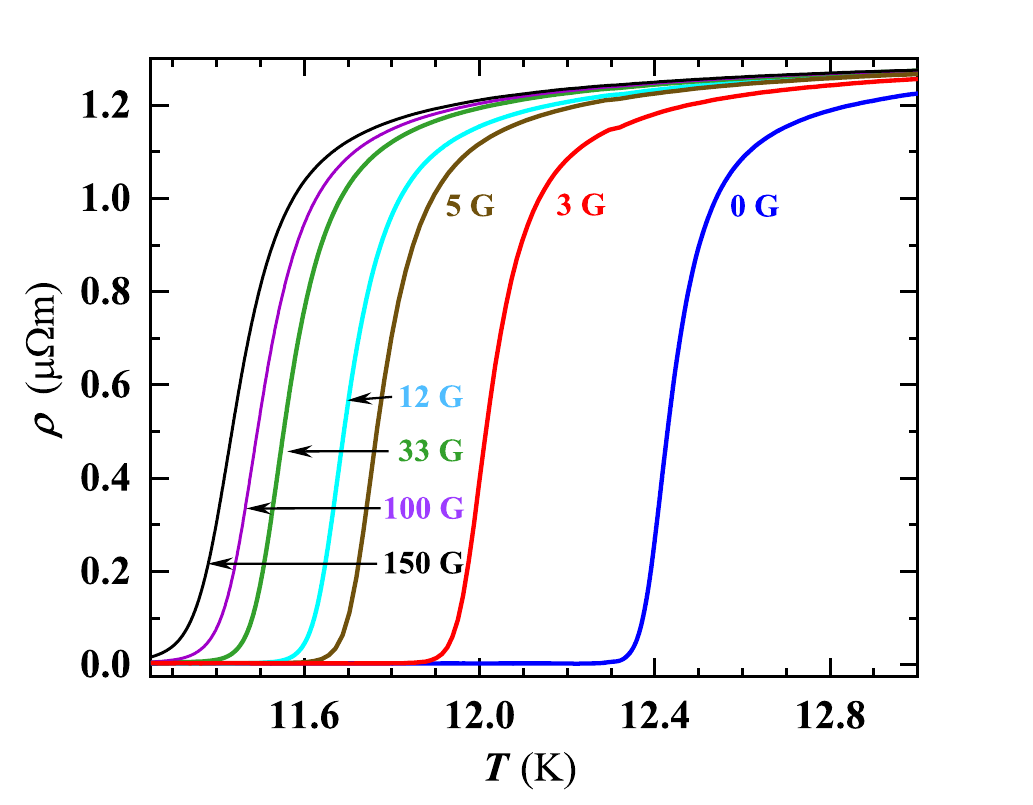}
\caption{Temperature dependent resistivity around the SC transition for a 10\,nm-thick NbN film 
(MO10) under several $H$ field intensities applied perpendicular to the film surface.}
\label{Res_vs_T}
\end{figure}
\begin{figure}[ht]
    \centering
    \includegraphics[width=0.55\textwidth]{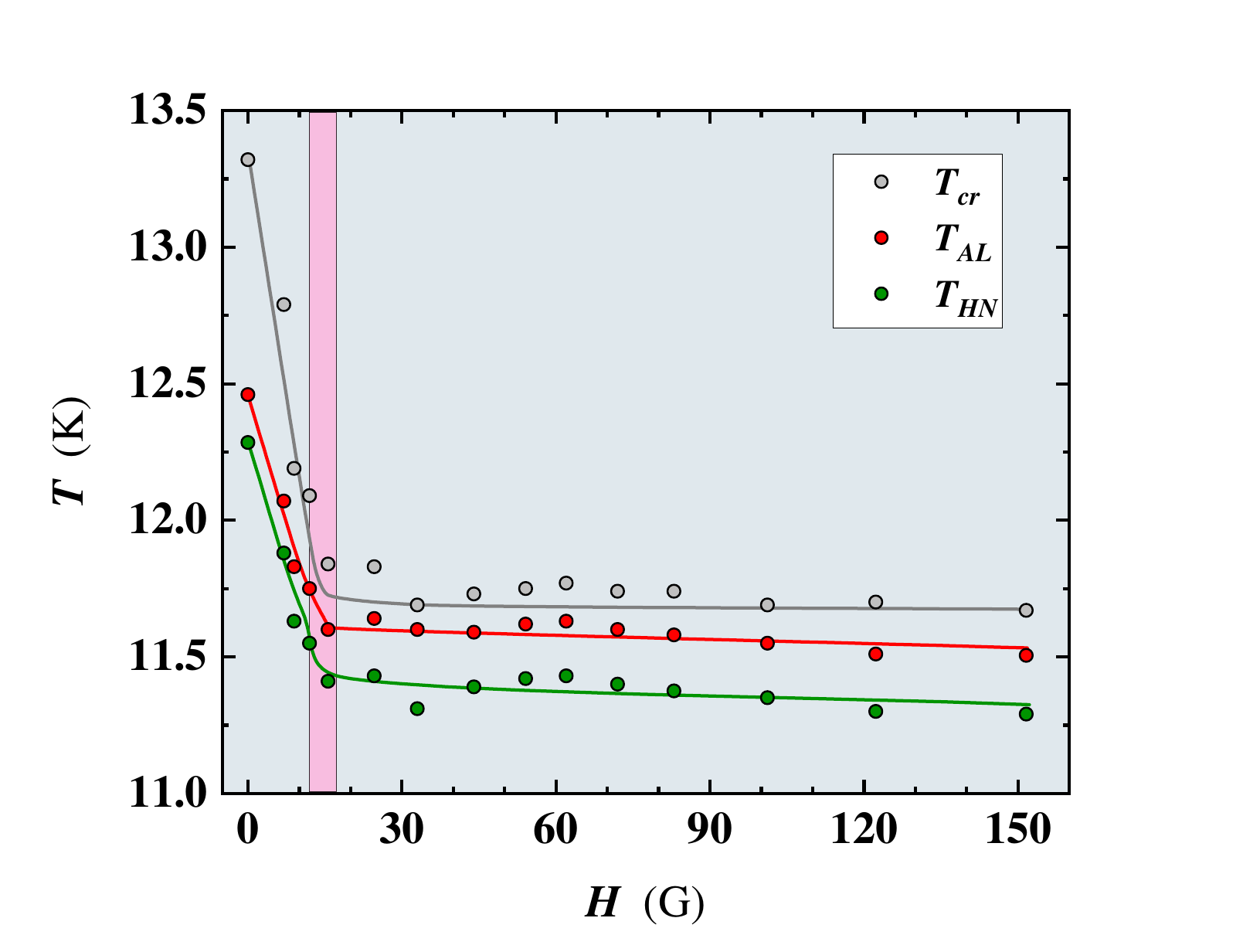}
    \caption{$H$ field intensity dependence of the characteristic temperature scales of film MO10, derived by: the crossing of the two fitting paraconductivity equations [(Eq.(1) and Eq.(2), in the main text] and the AL and HN critical temperatures. The lines are guides to the eye.}
    \label{Paracond}
\end{figure}
\subsubsection{Magnetic field dependence of the resistivity} 
We have explored the influence of very low $H$ fields ($3\div150$\,G) 
on the SC properties of studied NbN devices. This has been exploited  measuring the NbN resistivity, $\rho(T)$, near $T_\mathrm{C}$, under  diverse magnetic field intensities applied perpendicularly to the sample surface. Outcomes have revealed a pronounced downward shift of $T_\mathrm{C}$ for $H$ as low as 3\,G, which persists up to $\approx 33$\,G (Figure \ref{Res_vs_T}). 
Above this threshold, the shift in $T_\mathrm{C}$ progressively decreases. This trend has been attributed to the occurrence of a sizeable density of vortices, introduced by the applied field.
\subsubsection{Paraconductivity Analysis}
The analysis of the paraconductivity as a function of $H$ intensity by the Aslamazov-Larkin and Halperin-Nelson fitting equations [Eq.(1) and Eq.(2), respectively, in the main text] 
unambiguously evidences a clear trend in the $H$ field dependence of the temperature, $T_{cr}$, derived by the crossing of the two fittings equations. The value of $T_{cr}$ shows a rapid lowering, with the rise of the $H$ intensity, reaching a minimum at $\approx 15\div 20$ G then attaining an almost constant value (Figure~\ref{Paracond}).
\begin{figure}[h]
    \centering
    \includegraphics[width=0.5\textwidth]{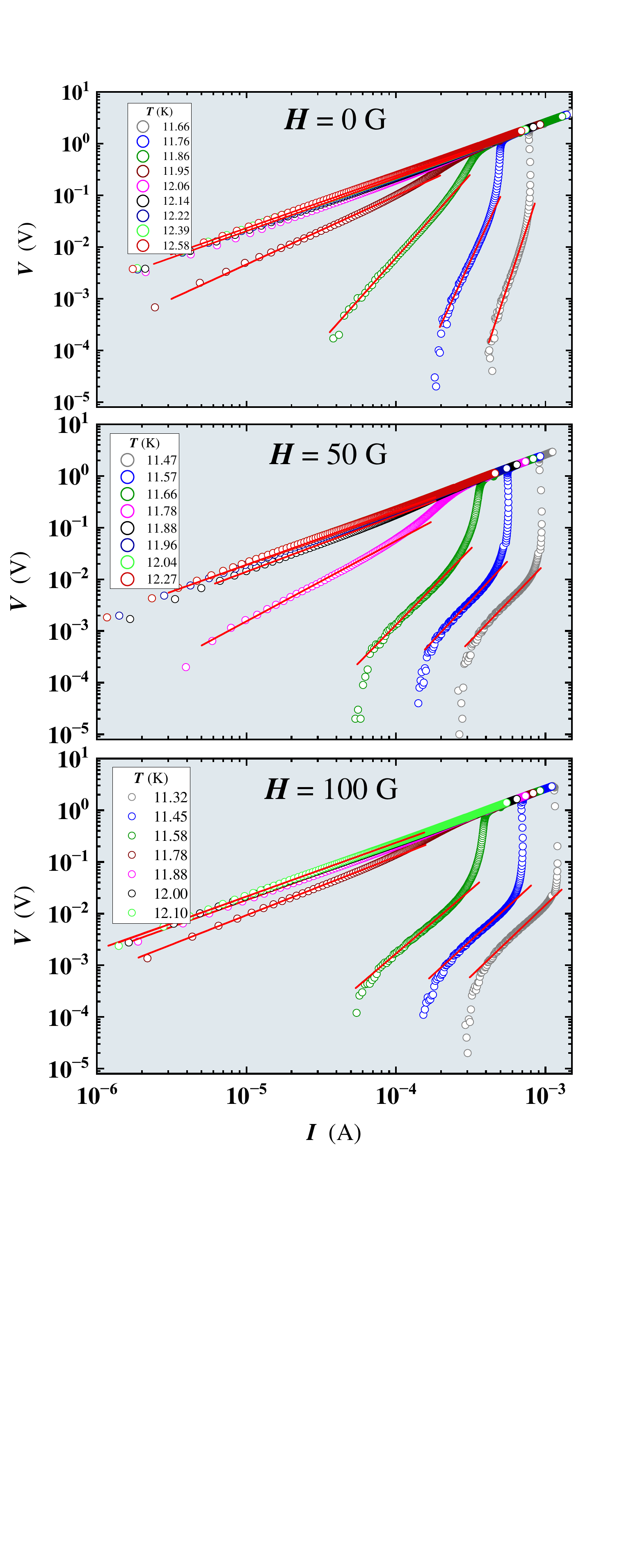}
    \vspace{-6cm}
    \caption{Temperature dependent $I-V$ characteristics of film MO10 under perpendicular applied magnetic field intensity of: $H = 0$ G, $H = 50$ G and $H = 100$ G. The red lines represent least squares fits by the relation $V\propto I(T)^{\alpha(T)}$ to extract the power law exponent $\alpha(T)$ at different $T$.}
    \label{Log_V-I}
\end{figure}
As discussed in the main text of the work, this finding is in agreement with the transition from a BKT to a BCS physics. A $H$ field dependent behavior similar to that found in $T_{cr}$ is observed in the critical temperatures $T_\text{AL}$ and $T_\text{HN}$, fitting parameters of the AL and HN equations, respectively (Figure~\ref{Paracond}). 
It's worthwhile mentioning that the fitting procedure has been carried on the same curve with two different equations and in two different $T$ regions. The similar $H$ field dependence of the accounted critical temperatures, confirm the validity and robustness of our analysis and conclusions.
\begin{figure}[h]
    \centering
    \includegraphics[width=0.50\textwidth]{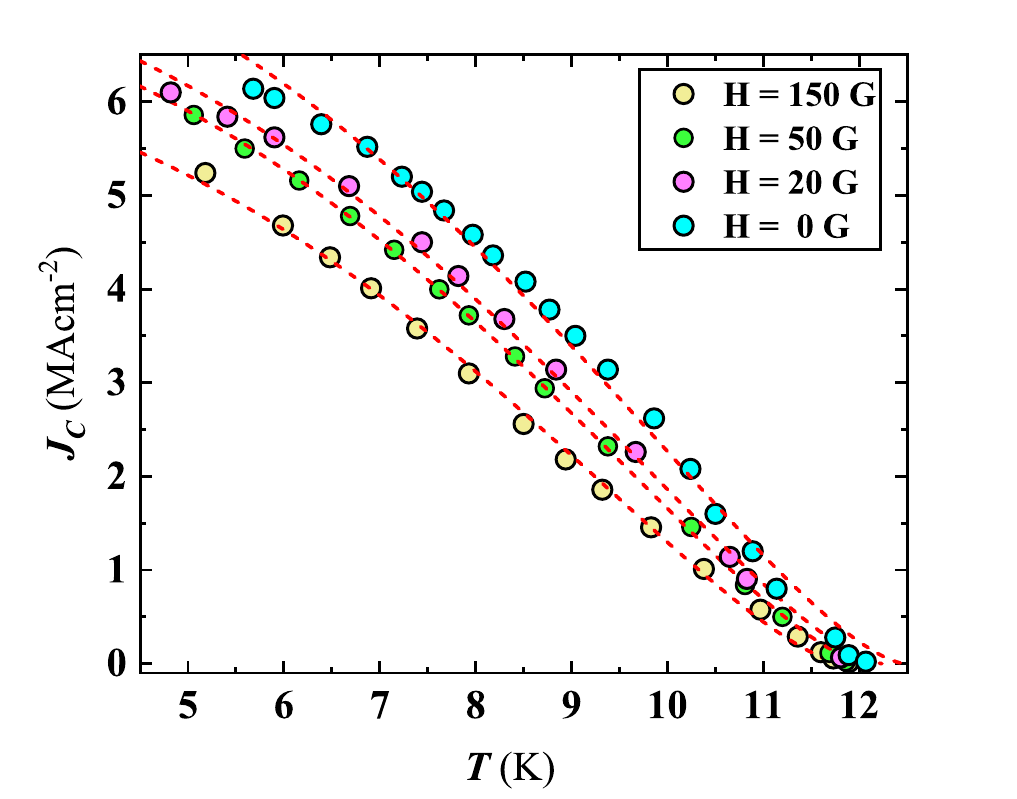}
    \caption{Temperature dependence of the superconducting critical current density, of film MO10, under several magnetic field intensities. The red broken lines are the least square fit of the experimental data points to the Ginzburg-Landau equation (Eq.~\ref{Jc_vs_T}).}
    \label{J_C0}
\end{figure}
\subsubsection{$I-V$ characteristics}
The temperature dependence of the superconducting critical current density, $J_\mathrm{C}(T)$, shown in the Figure 2 of the main text, has been derived by measured $I-V(T)$ curves. The value of the superconducting critical current, $I_\mathrm{C}(T)$, has been defined as the current intensity at which a maximum in the $dV(I)/dI$ curve occurs.\\
The value of the superconducting critical current density at 0 K, $J_\mathrm{C0}$, has been extracted by a least-squares fit of the $J_\mathrm{C}(T)$ plot (see Figure~\ref{J_C0}), according to the Ginzburg-Landau relation:
\begin{equation}
 J_\mathrm{C}(T) = J_\mathrm{C0}\left[1-\left(\frac{T}{T_\mathrm{C}}\right)^{2}\right]^\frac{3}{2}\left[1+\left(\frac{T}{T_\mathrm{C}}\right)^{2}\right]^\frac{1}{2}
\label{Jc_vs_T}
\end{equation}

\subsection{Superfluid Stiffness}
The superfluid stiffness $J_s(T)$ and the exponent $\alpha(T)$ of the power-law relation $V\propto I(T)^{\alpha(T)}$ have been extracted by a least squares fitting of $V-I(T)$ curves measured around $T_\mathrm{BKT}$, as reported in the main text (see also Figure 6). As mentioned above, a pulsed technique has been used to measure the $I-V$ curves. By employing this method, potential artifacts associated with the heating of the device under test during the complete measurement are minimized. This fact has allowed to extend the fitting region of the sourced current up to $\lesssim 1$ mA (Figure~\ref{Log_V-I}).
\bibliographystyle{apsrev4-2}  
\bibliography{PRL}  
\end{document}